\documentclass[11pt]{article}
\usepackage{moriond,epsfig,subfigure,amssymb}

\def\be{\begin{equation}}
\def\ee{\end{equation}}
\def\bea{\begin{eqnarray}}
\def\eea{\end{eqnarray}}

\begin{document}
\vspace*{4cm}
\title{RECENT RESULTS ON HIGH-$Q^2$ NEUTRAL \\
       AND CHARGED CURRENT CROSS SECTIONS AT HERA}
\author{Takahiro Fusayasu}
\address{Department of Physics, Graduate School of Science,\\
University of Tokyo, Hongo 7-3-1, Bunkyo-ku, Tokyo 113-0033, Japan\\
{\rm (On behalf of the H1 and ZEUS Collaborations)}}

\maketitle
\abstracts{
  High-$Q^2$ NC and CC DIS cross sections have been measured by
  H1 and ZEUS at HERA.
  Both NC and CC results based on data taken during the year 1994-2000 are
  in good agreement with Standard Model expectations.
  The structure function $xF_3^{NC}$ is extracted from the NC cross sections
  and the mass of the $W^\pm$ propagator is extracted from
  the CC cross sections.
  Valence quark distributions are derived by means of an NLO QCD fit.
}

\section{Deep Inelastic Scattering at HERA}
HERA collides a positron or electron beam with a proton beam.
From 1994 to 1997 HERA operated as an $e^+p$ collider with 
a centre-of-mass energy $\sqrt{s}=300$~GeV.
In 1998 it started in the $e^-p$ mode with $\sqrt{s}$ raised to 318~GeV,
and then switched to $e^+p$ at the same energy in July 1999.
The H1 and ZEUS experiments have collected 102 and 114 pb$^{-1}$ of $e^+p$
and 16.4 and 16.7 pb$^{-1}$ of $e^-p$ integrated luminosity, respectively.
The available $Q^2$ of the experiments can reach as high as $10^5$ GeV$^2$,
which corresponds to a spatial resolution of about 10$^{-3}$ fm.
This paper presents results on neutral (NC, $e^{\pm}p \to e^{\pm}X$)
and charged current (CC,
$e^{\pm}p \to \raisebox{5pt}{\tiny(}\bar{\nu}\raisebox{5pt}{\tiny)} X$)
high-$Q^2$ deep inelastic scattering (DIS)
based on 81.5~(47.7)~pb$^{-1}$ of H1~(ZEUS) $e^+p$ data%
~\cite{h1:ccnc9497,zeus:nc9497,zeus:cc9497,h1:ccnc9900}
and on the full luminosity of the $e^-p$
data~\cite{h1:cc9899,zeus:nc9899,zeus:cc9899}.

The Born cross section for NC DIS is written as:
\begin{equation}
\label{eq:ncred}
\frac{d^2\sigma_{NC}^{e^{\pm}p}}{dx dQ^2} =
  \frac{2 \pi \alpha^2}{xQ^4}
  \left[
         Y_{+}F_2^{NC}(x,Q^2)
     \mp Y_{-}xF_3^{NC}(x,Q^2)
      -  y^{2}F_{L}^{NC}(x,Q^2)
  \right]
\end{equation}
where $\alpha$ denotes the electromagnetic fine structure constant,
$x$ the Bjorken scaling variable, $y$ the elasticity parameter
and $Y_{\pm}=1\pm(1-y)^2$.
The reduced cross section, $\tilde{\sigma}_{NC}$, is often used,
and is defined as the double differential cross section of Eq.~(\ref{eq:ncred})
divided by the factor $(\frac{2 \pi \alpha^2}{xQ^4})Y_+$,
so that the structure function component is extracted.
The CC DIS cross section is expressed at leading-order in QCD as:
\begin{eqnarray}
\frac{d^2\sigma_{CC}^{e^+p}}{dx dQ^2} &=&
  \frac{G_F^2}{2 \pi}
  \left(\frac{M_W^2}{M_W^2 + Q^2}\right)^2\,
  \sum_{i=1}^2\,\left[\overline{u}_i(x,Q^2) + (1-y)^2\,d_i(x,Q^2)\right]
  \label{eq:ccred_pos}\\
\frac{d^2\sigma_{CC}^{e^-p}}{dx dQ^2} &=&
  \frac{G_F^2}{2 \pi}
  \left(\frac{M_W^2}{M_W^2 + Q^2}\right)^2\,
  \sum_{i=1}^2\,\left[u_i(x,Q^2) + (1-y)^2\,\overline{d}_i(x,Q^2)\right]
  \label{eq:ccred_ele}
\end{eqnarray}
where $G_F$ denotes the Fermi coupling constant
and $M_W$ the mass of the $W^\pm$ boson.
The $u_i$ and $d_i$ are $u$- and $d$-type quark densities in the proton,
respectively, with $i$ running over the first and second generations.
As in the NC case,
the CC reduced cross section, $\tilde{\sigma}_{CC}$, is defined as
Eq.~(\ref{eq:ccred_pos}) or Eq.~(\ref{eq:ccred_ele})
divided by the factor
$\frac{G_F^2}{2 \pi x}(\frac{M_W^2}{M_W^2 + Q^2})^2$.

\section{Cross Section Measurement}
\begin{figure}
  \begin{center}
  \psfig{figure=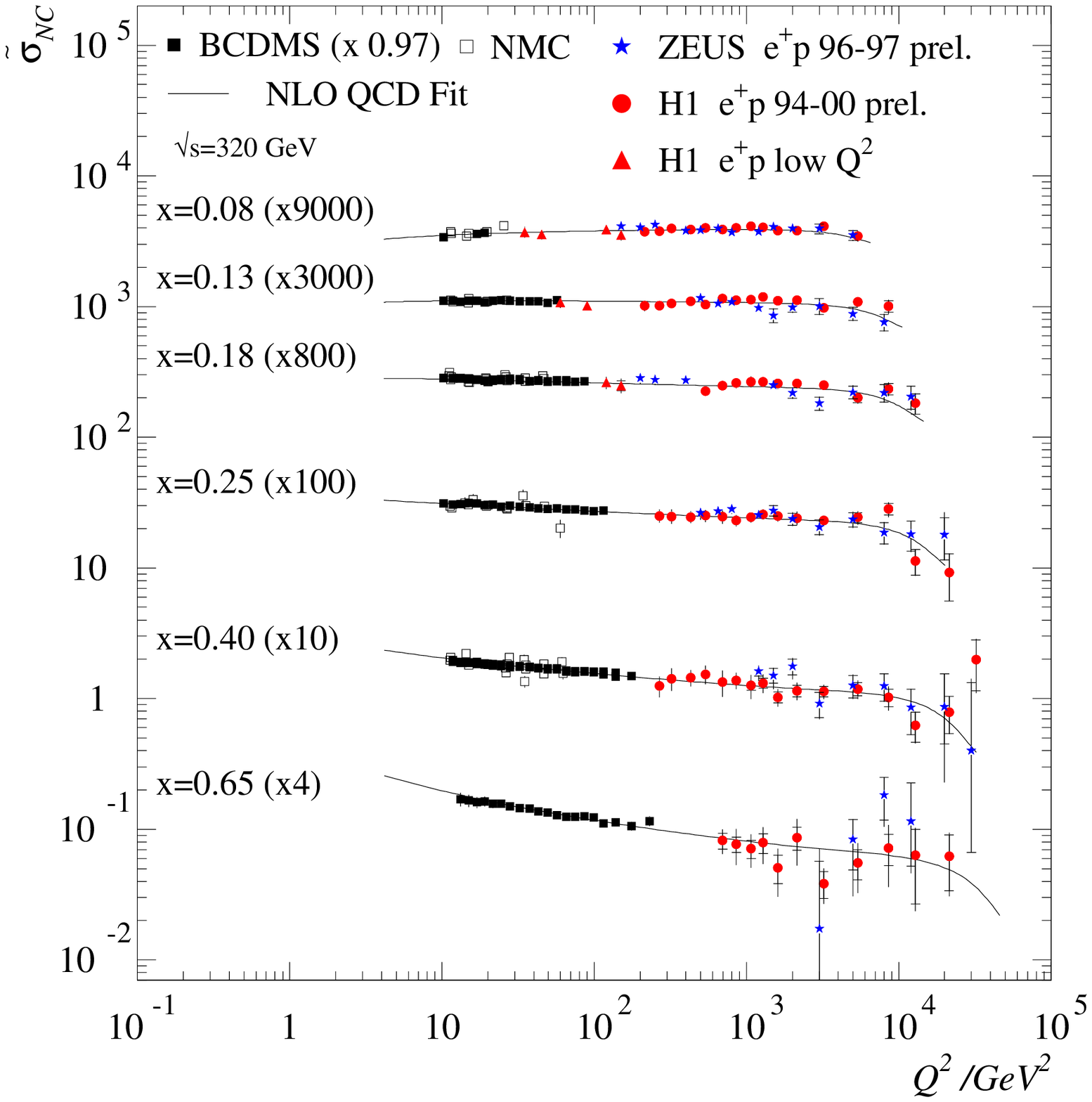,height=7.8cm}
  \psfig{figure=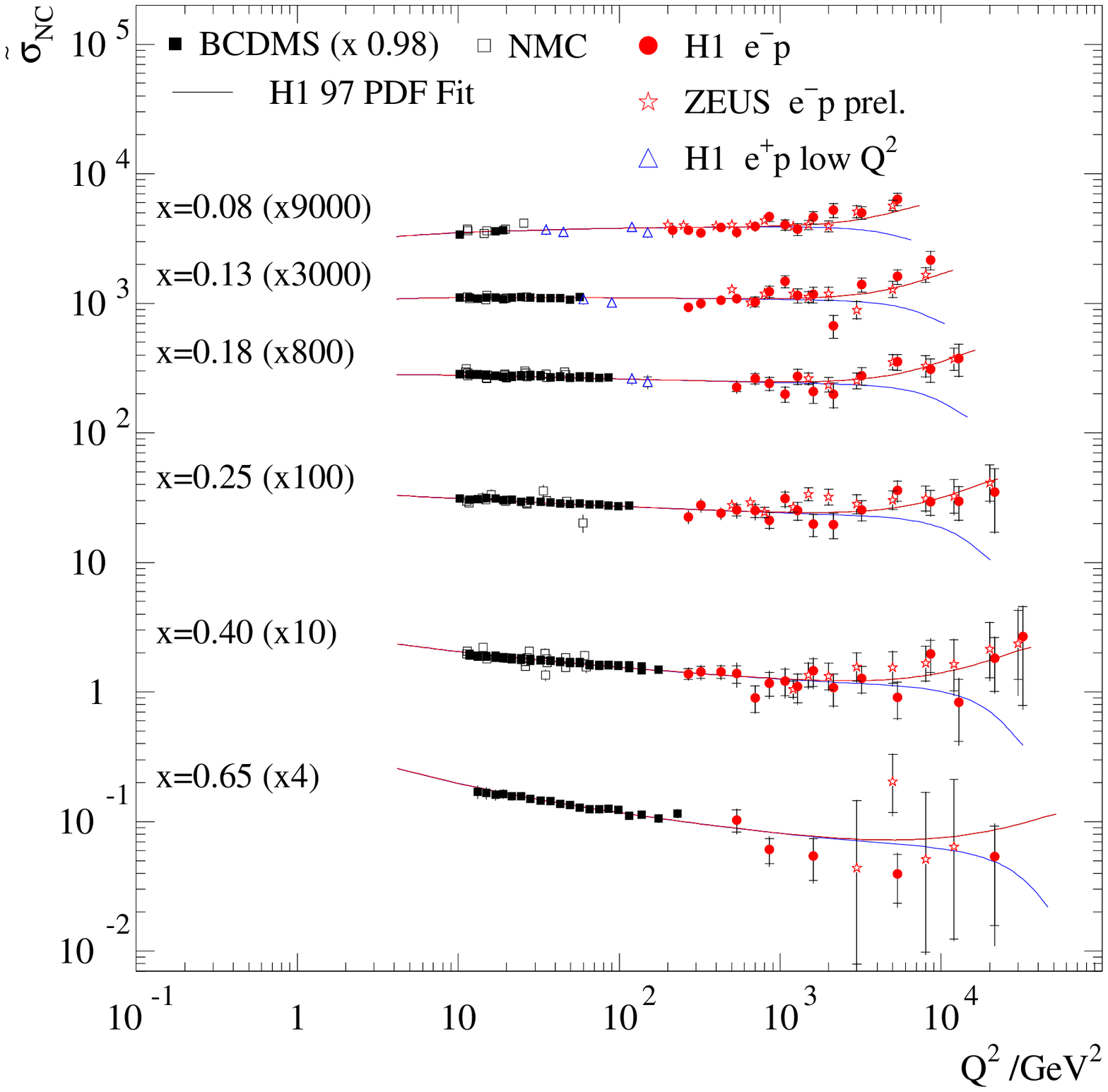,height=7.8cm}
  \end{center}
  \caption{The reduced NC cross sections, $\tilde{\sigma}_{NC}$, compared
           to the SM predictions for $e^+p$ (left) and $e^-p$ (right).}
  \label{fig:ncddq}
\end{figure}

The NC and CC DIS cross sections are both measured for $Q^2>200$ GeV$^2$.
Figure~\ref{fig:ncddq} shows the NC reduced cross sections
$\tilde{\sigma}_{NC}$ as a function of $Q^2$ for fixed $x$ values.
The inner and outer error bars represent
the statistical and total uncertainties, respectively.
Also shown are the standard model (SM) expectations based on PDFs
obtained from an NLO QCD fit~\cite{h1:ccnc9497}.
The results are in good agreement with SM predictions over the kinematic range.
This illustrates the successful description of the NLO QCD evolution
with the DGLAP equation up to $Q^2\sim30,000$ GeV$^2$.
Due to the $\gamma$-$Z^0$ interference term, $xF_3^{NC}$,
we observe small differences between the $e^+p$ and $e^-p$
cross sections at high $Q^2$.
We discuss $xF_3^{NC}$ in detail in Section~\ref{sec:others}.

\begin{figure}
  \begin{center}
  \psfig{figure=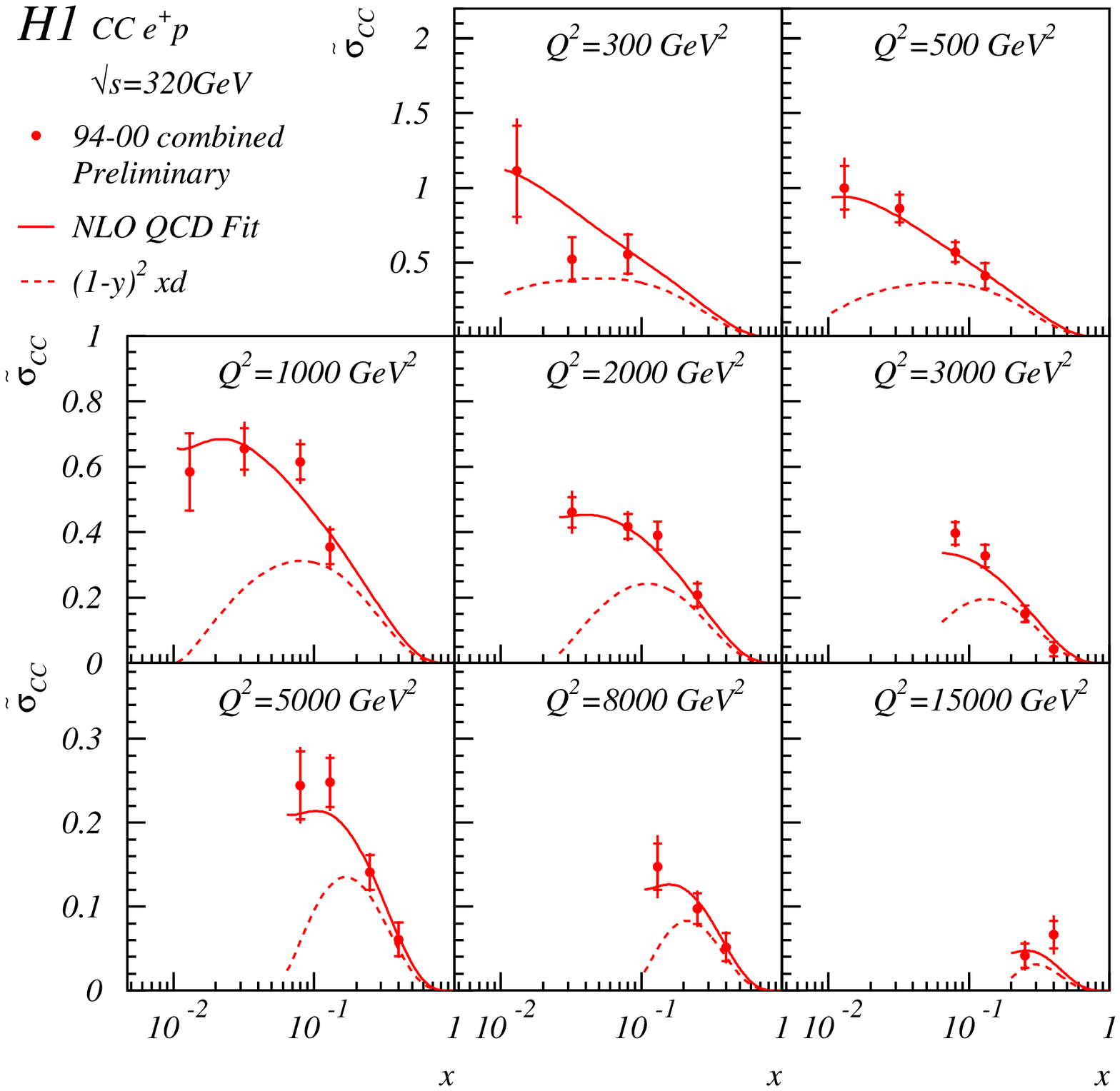,width=7.2cm}
  \psfig{figure=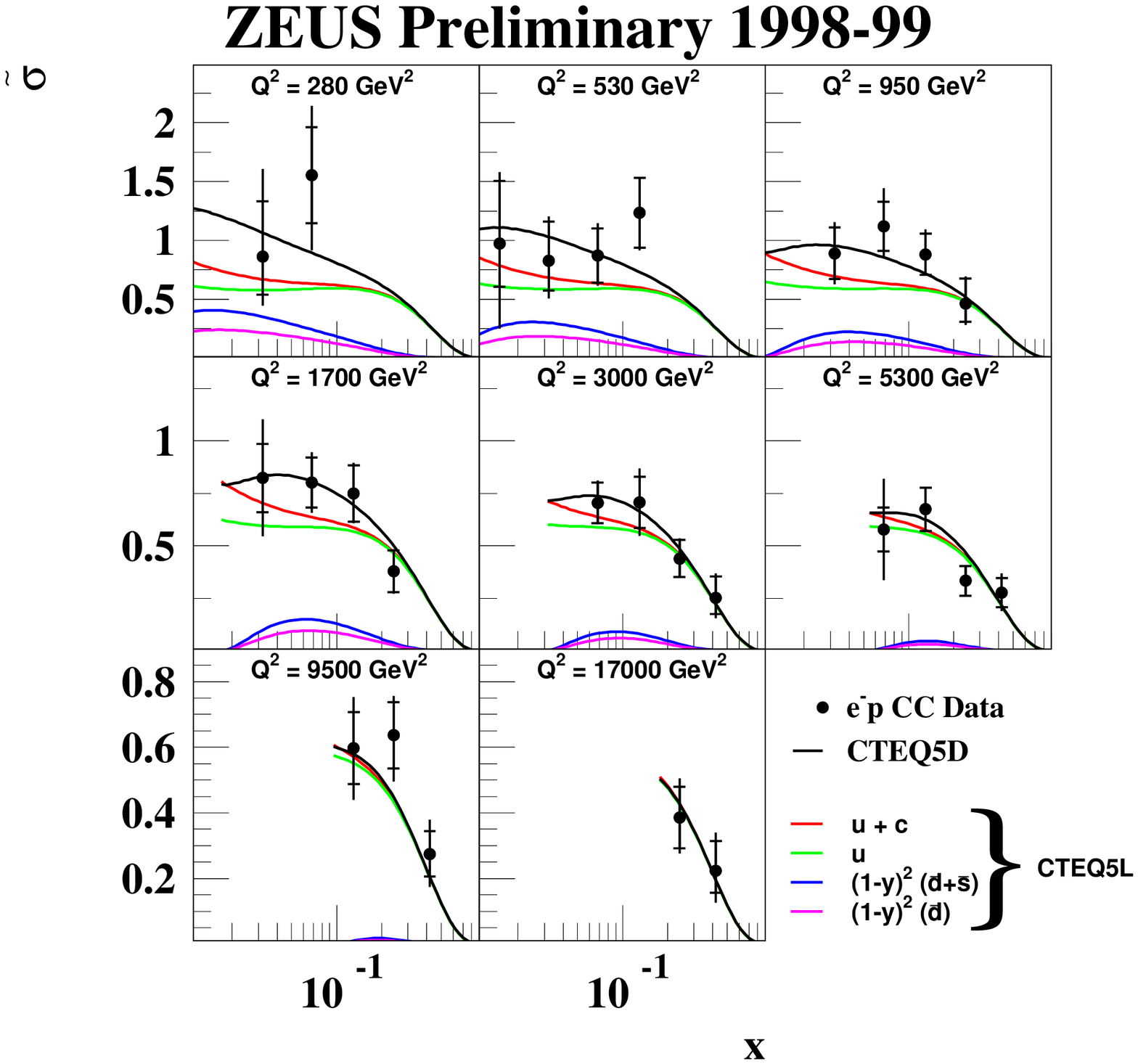,width=7.8cm}
  \end{center}
  \caption{The reduced CC cross sections, $\tilde{\sigma}_{CC}$, compared
           to the SM prediction for $e^+p$ (left) and $e^-p$ (right).}
  \label{fig:ccddx}
  \vspace{-4pt}
\end{figure}

Figure~\ref{fig:ccddx} shows the CC reduced cross sections,
$\tilde{\sigma}_{CC}$, as a function of $x$ for fixed $Q^2$ values,
together with the SM predictions based on the NLO QCD fit~\cite{h1:ccnc9497}
(left plot) or CTEQ5~\cite{cteq5} (right plot).
The measured cross sections are consistent with the SM predictions.
The expected contribution from each quark density is also shown.
This illustrates that the CC cross section at high $x$
in $e^+p$ collisions is dominated by the $d$ quark,
while in $e^-p$ collisions it is dominated by the $u$ quark.

\section{Extraction of $xF_3^{NC}$, $M_W$ and
         Valence Quark Distributions}\label{sec:others}
\begin{figure}
  \begin{center}
    \begin{tabular}{cc}
      \subfigure[]{\psfig{figure=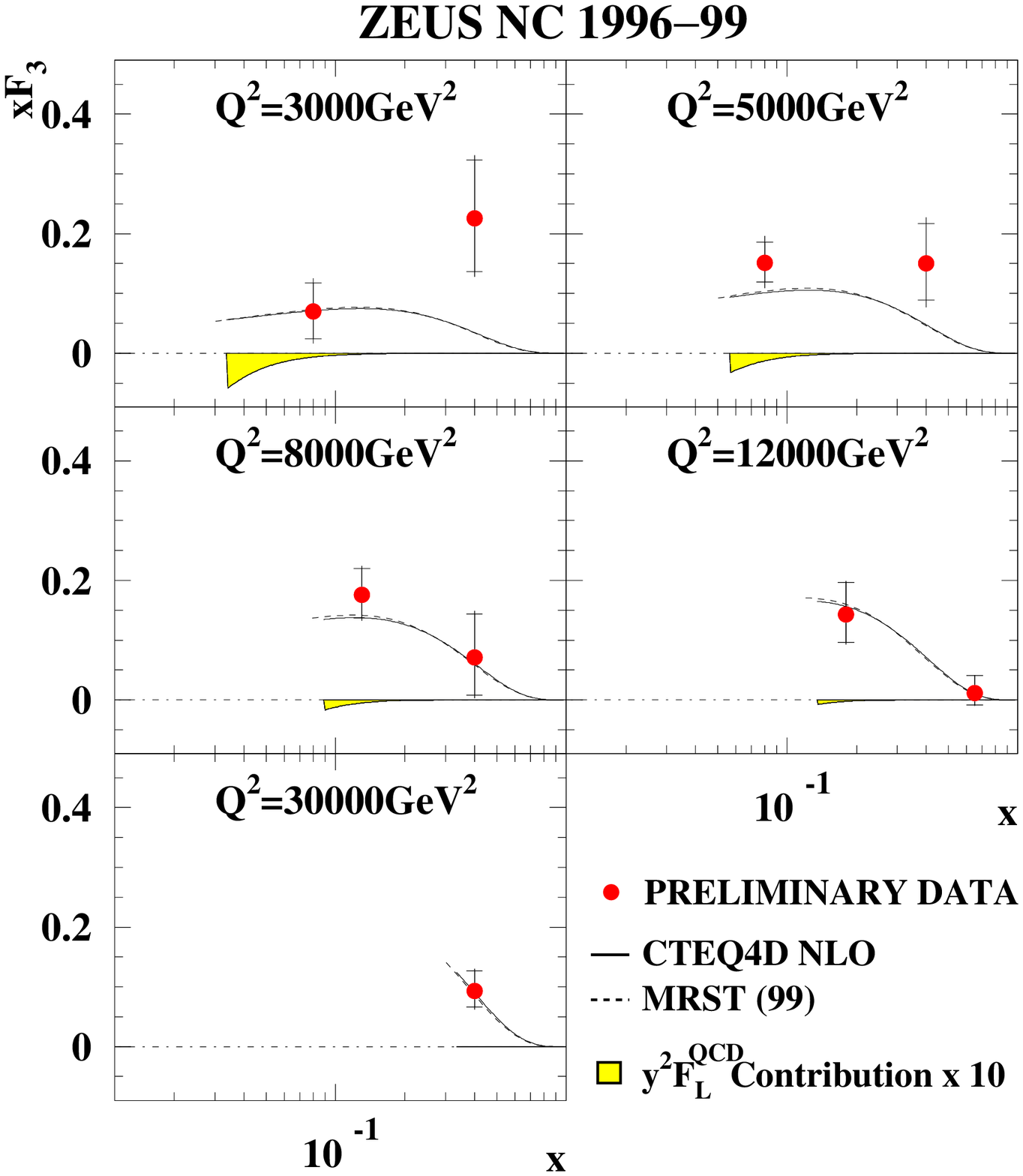,height=8.0cm}\hskip 5mm}
      \subfigure[]{\psfig{figure=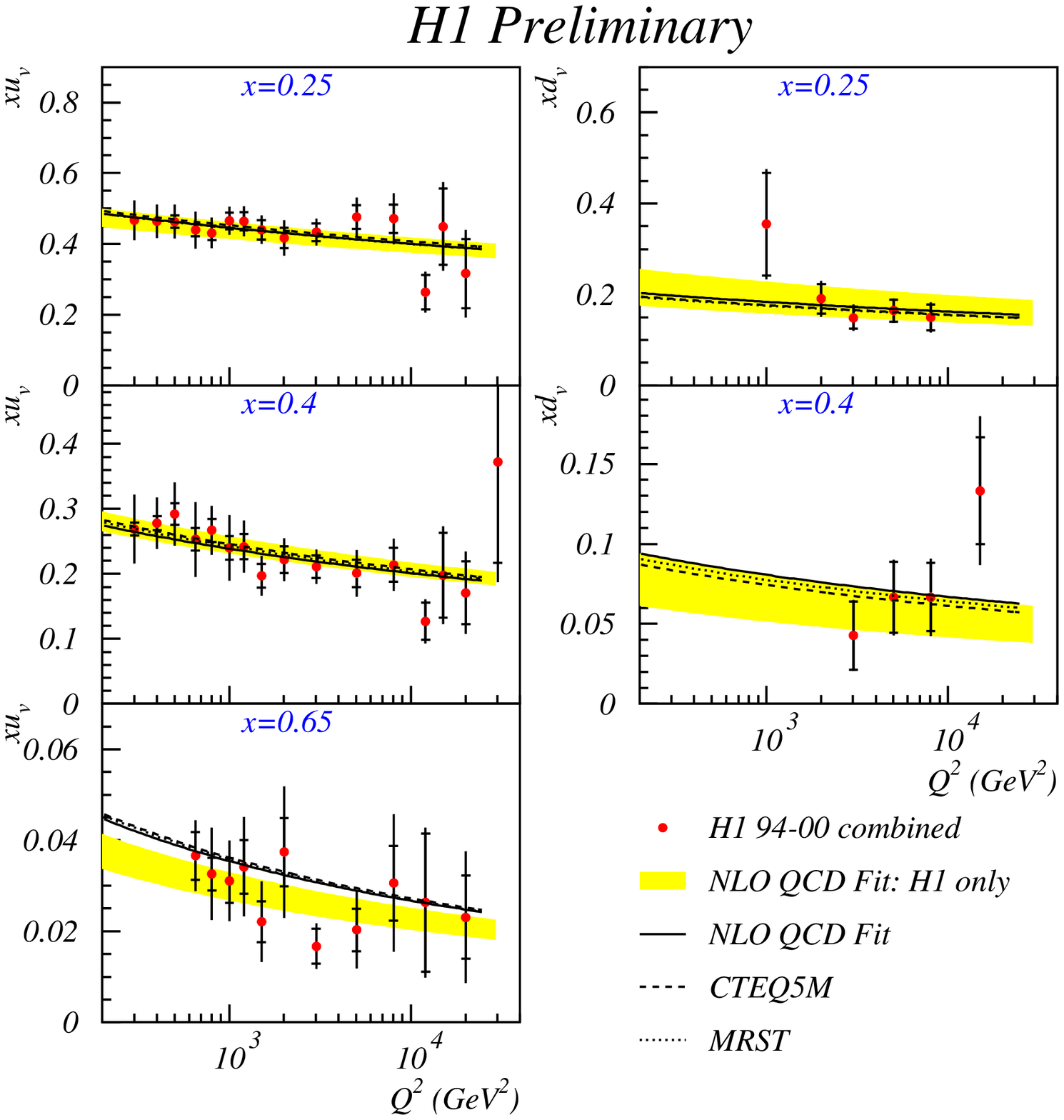,height=8.0cm}}
    \end{tabular}
  \end{center}
  \vspace{-10pt}
  \caption{(a) $xF_3^{NC}$ extracted from the NC double differential
               cross sections.
           (b) Valence quark densities derived from the NC and CC
               cross sections by means of an NLO QCD fit.}
  \label{fig:xf3_and_qval}
\end{figure}

Using the NC double differential cross sections from
the 1994-1997 $e^+p$ data at $\sqrt{s}=300$ GeV
and the 1998-1999 $e^-p$ data at $\sqrt{s}=318$ GeV,
$xF_3$ is determined from:
\begin{equation}
  xF_3^{NC} = \frac{xQ^4}{2 \pi \alpha^2} \cdot
       \left(
         \frac{Y_{-}^{300}}{Y_{+}^{300}}+\frac{Y_{-}^{318}}{Y_{+}^{318}}
       \right)^{-1} \cdot
       \left(
         \frac{1}{Y_{+}^{318}}\frac{d^2\sigma_{NC}^{e^-p}}{dx dQ^2} -
         \frac{1}{Y_{+}^{300}}\frac{d^2\sigma_{NC}^{e^+p}}{dx dQ^2}
       \right) - \Delta F_L
\end{equation}
where ``300'' and ``318'' denote the different $Y_\pm$ obtained using
the different centre-of-mass energies, $\sqrt{s}$.
$\Delta F_L$ is a correction due to the longitudinal structure function, $F_L$.
The size of $\Delta F_L$ obtained by QCD calculations
is less than 1\% over the kinematic ranges used in this paper,
and becomes as large as 10\% in the lowest $x$ and $Q^2$ region.
Figure~\ref{fig:xf3_and_qval}(a) shows
the results from ZEUS~\cite{zeus:nc9899}.
To reduce the statistical fluctuations, several bins of
the double differential measurements are combined.
The size of the $\Delta F_L$ effect is shown by the shaded bands
and is negligible when compared to the size of the statistical uncertainty.

$M_W$ is extracted by a $\chi^2$-fit to the CC $d\sigma/dQ^2$
based on ZEUS 1994-1997 $e^+p$ data with $M_W$ as a free parameter.
The fit yielded
$M_W = 81.4 ^{+2.7}_{-2.6}({\rm stat.})
   \pm 2.0({\rm syst.}) ^{+3.3}_{-3.0}({\rm PDF})$ GeV.
H1 has applied a similar fit to their CC double differential
cross sections and obtained
$M_W = 81.2 \pm 3.3({\rm stat.}) \pm 1.7({\rm syst.}) \pm 3.7({\rm PDF})$~GeV
from the 1994-1997 $e^+p$ data and
$M_W = 79.9 \pm 2.2({\rm stat.}) \pm 0.9({\rm syst.}) \pm 2.1({\rm PDF})$~GeV
from the 1998-1999 $e^-p$ data.
These space-like measurements are in good agreement with
the world average of time-like measurements.

H1 has used~\cite{h1:ccnc9900} their combined $e^+p$ double differential
NC and CC cross sections together with the $e^-p$ data in an NLO QCD fit
to determine the dominant valence quark distributions $xu_v$ and $xd_v$
at high $Q^2$ and high $x$.
The H1 low-$Q^2$ data~\cite{h1:lowq2} are also included
in order to constrain the gluon and sea quarks.
The results are shown as the shaded error bands
in Figure~\ref{fig:xf3_and_qval}(b)
together with other parametrisations from MRST~\cite{mrst}, CTEQ5~\cite{cteq5}
and the previous H1 fit~\cite{h1:ccnc9497}.
Also shown are the valence quark densities determined with the
local extraction method defined as:
\begin{equation}
  xq_v(x,Q^2) = \sigma_{meas}(x,Q^2)
                \left(\frac{xq_v(x,Q^2)}{\sigma(x,Q^2)}\right)_{para}
\end{equation}
where the first factor $\sigma_{meas}(x,Q^2)$ on the right-hand-side
is the measured NC or CC double differential cross section,
and the second factor is the theoretical expectation
from the H1 fit~\cite{h1:ccnc9497}.

\section{Summary}
NC and CC DIS cross sections in $e^+p$ and $e^-p$ collisions
measured by the H1 and ZEUS experiments
are in good agreement with SM predictions over the kinematic range of
$200 \lesssim Q^2 \lesssim 30000$ GeV$^2$ and $0.01 \lesssim x \lesssim 0.65$.
$xF_3^{NC}$ is determined from the NC double differential cross sections.
$M_W$ has been extracted from the CC cross sections
and is found to be consistent with LEP and Tevatron measurements.
The valence quark distributions have been extracted by means of an NLO QCD fit
to the NC and CC double differential cross sections.

\section*{Acknowledgements}
I would like to thank H1 and ZEUS members for helpful discussions.
I also thank the organisers for inviting me to the excellent conference.
The author has been supported by the Japan Scholarship Foundation.

\section*{References}

\end{document}